\documentclass[12pt]{article}
\usepackage{graphicx}
\usepackage{color}
\begin{document}

\title{On the choice of ingredients for a theory of the Ice Ages} 

\author{
Walter Baltensperger\footnote{Centro Brasileiro de Pesquisas F\'\i sicas,  Rio de Janeiro, Brazil;  e-mail: baltens@phys.ethz.ch}\ \ and Willy Woelfli\footnote{ Institute for Particle Physics, ETHZ H\"onggerberg,  Z\"urich, Switzerland (Prof.~emerit.).} }
\date{}
\maketitle
\begin{abstract} \noindent
``With five parameters one can fit an elephant.'' This provocative statement expresses the fact that when a theory has  several adjustable parameters, an agreement with empirical data can be of modest value. What about a theory which contains unobserved objects? This is the subject of this paper. It is motivated by a model of the Ice Ages of the  Pleistocene, which postulates a hot planet in an extremely eccentric orbit. This object has many consequences. It is rather well defined by the requirements, that it must not be in conflict with laws of nature, nor with empirical data. It must have sufficient mass  to produce a rapid geographic  pole shift on Earth after a close flyby at the end of the Pleistocene, and also be small enough to disintegrate at this occasion and to evaporate during the Holocene. These requirements leave hardly any adaptable parameters.
In this situation, the agreement with further data, in particular the reverse Dansgaard-Oeschger events of the Holocene,  represents a significant support of this theory.
\end{abstract}
%\tableofcontents
\section{Ingredients of a model}
\subsection{Laws of nature}

Which ingredients belong to a scientific theory of a complex phenomenon? First,  of course, the laws of nature. They must not be violated by the model.  Often a rigorous treatment of the model is impossible. In this case it should be plausible that the approximations do not dominate the result. For a mathematician, who requires rigor, this can be a difficult task.  Most scientists will be satisfied, when the neglected terms appear to be relatively small.
\subsection{The  obvious objects of the system}
The objects which obviously belong to the system must be part of the model. When climate is studied, the shape of  oceans and continents, the properties of water and air and the basic  parameters of the Sun evidently belong to the model. In studies involving prolonged times, the other planets and their gravitational coupling to Earth must be included.

\section{The case of the Milankovitch theory}
\subsection{Clearly correct ingredients of the model}
The  Pleistocene Ice Ages lasted from about 3\thinspace 000 ka  ago, where  1 ka = 1\thinspace 000 years, \cite[fig.1]{Woelfli2011}\footnote{Frequently figures from ref.\cite{Woelfli2011} will be cited, where the origin of the data is quoted.}
until about 10 ka  ago \cite[fig.3]{Woelfli2011}. 
The usual theory of this cold climate starts with clearly correct ingredients. The
force of gravity between Earth and the other planets lead to variations of Earth's orbit, which have first been described by Milankovitch \cite{Milankovitch1941}. The orbital parameters of Earth vary slowly with approximate frequencies. 
These changes follow from the laws of mechanics and gravity and can be calculated in detail \cite{Berger}. The Milankovitch variations of Earth's orbital parameters are not questioned. They belong to any theory of the Ice Ages. 

\subsection{Are these ingredients enough?}
When the Ice Ages started about 3\thinspace 200 ka  ago, the climate became colder on the average, while
  global variations of the temperature increased
 \cite[fig.1]{Woelfli2011}.  About 2\thinspace 000 ka  ago the Ice Ages were in full swing. About 10 ka ago a new period started, which was about 3$^\circ$ C warmer than at present  \cite[figs.\thinspace 3,10,11]{Woelfli2011}: the Holocene. Now, we are privileged to live in the calm climate of the last 300 years. 

The usual theory studies amplification mechanisms, by which the changes of insolation due to the variations of Earth's orbital parameters may produce the observed large modifications of the climate. The climate involves many feedbacks. Positive feedbacks lead to amplifications. As an example, when it is cold, snow appears, which reflects more sunlight back to outer space.  This further lowers the temperature. Positive feedbacks destabilize the system. When oceans become warmer, more water evaporates, so that more white clouds appear. The increased reflection of sunlight lowers the temperature. This feedback is negative. It stabilizes the climate. On the other hand, warmer oceans release more carbon dioxide and water vapor into the atmosphere. This leads to a positive feedback. 

The reflection of sunlight depends on the vegetation on land, on the color of water and even on the agitation of its surface. Other complex feedbacks are due to changes of currents of the oceans and of the atmosphere.  The feedbacks present a complex problem, practically impossible to quantify from first principles. In this situation, feedbacks become parameters that can be fitted to some extend.

Computer programs for weather or  climate contain feedback mechanisms.  Corresponding parameters can be adjusted in  the programs for best results with data of the current weather or climate. In a way, these adjusted values are then  measured values. The present climate appears to be quite stable, since small variations of the parameters lead to small changes of the climate. 

\subsection{The non linear theory}
To obtain a theory of the Ice Ages, it is necessary to assume that these coefficients are sufficiently strong functions of temperature, humidity etc., so that the climate can enter into a completely different mode. The properties of the resulting climate  are not linear extrapolations of changes obtained with small variations of the established coefficients. For a description of the Ice Ages a non-linear theory is required. When these increased feedbacks are adjusted to the unusual phenomenon,  the criticism that ``with five parameters one can fit an elephant" becomes relevant. 

In the non-linear theory the Ice Ages are separated from the normal climate. It is not apparent, how the small variations of Earth's orbital parameters induce a passage from one to the other.  Usually, the question how the Ice Ages began is not considered. The Holocene including the present time is assumed to be  a warm interval of the Ice Ages.

\subsection{Success and failure}
The Ice Ages contain at least one feature, which is not likely to be explained by the non-linear theory: the asymmetry of the glaciation with respect to the (present) North Pole \cite[fig.\thinspace 5, 6]{Woelfli2011}. 
At the time of maximum glaciation, about 20 ka ago, the ice reached the region where New York City stands at present, while East Siberia remained ice free.
Mammoths, which feed on steppe grass, lived in Arctic regions of East 
Siberia \cite[subsec.\thinspace 2.5]{Woelfli2011}. If geographically
these regions had been Arctic at that time, then the yearly insolation would have been insufficient for the growth of steppe grass.

The Earth is approximately a sphere and Earth's orbit nearly a circle with almost constant radius. Therefore, the total solar radiation that falls on Earth is nearly constant.  The obliquity, i.e. the angle between Earth's axis and the perpendicular to its orbital plane, creates the seasons during the yearly orbit around the Sun.. The larger the obliquity, the more summer differs from winter on both hemispheres. The obliquity varies between 22.1$^\circ$ and 24.5$^\circ$  with an approximate 41 ka Milankovitch period.  To the extend to which warmer summers are compensated by colder winters, a slight increase of the obliquity does not modify the average global temperature. To change this, a non linear response is required.  Similar  situations appear also with the variations of other orbital parameters.

Ice bores in Greenland and Antartica and sediments of the ocean bottom have revealed the climate of past times. Certain Milankovitch periods have been experimentally uncovered. These are truly great scientific achievements of the 20$^{\rm th}$ century. 

However, the agreement with the
Milankovitch  theory shows failures. During the last 800 ka, the most important period of the climate was 100 ka \cite[fig.15]{Woelfli2011}. This somewhat fits with  calculated periods of the ellipticity of Earth's orbit in the range from 95 ka to 125 ka. However, the orbital ellipticity is small, so that only a minor effect on climate is expected. If somehow this was sufficiently amplified, then the rather strong beat period of about 400 ka should also appear, which is not visible in the climate data. On the other hand, 
between 1\thinspace 200 ka  and 800 ka ago the dominant temperature variation agrees with the 41 ka variation of the obliquity.

The most violent temperature variations of the Ice Ages, the Dansgaard-Oeschger variations,  were quite irregular and much too fast to be connected with Milankovitch periods \cite[fig.\thinspace 8]{Woelfli2011}. In the usual theory they are mostly attributed to instabilities of currents of the oceans and the atmosphere.

\section{The case of the hot planet}
\subsection{The problem of unobserved objects}

Clearly,  when a theory has many parameters to fit,  then an agreement with measured data represents modest support. What about a theory with unobserved  objects, which have been introduced to explain the data? This is the subject of this paper. Is the introduction of an unobserved object into a theory not equivalent to introducing many adaptable parameters, so that an agreement with measured data is not surprising and of little value?

As an example of an unobserved object consider the asteroid,  which supposedly
caused  the extinction of the dinosaurs about 60\thinspace 000 ka  ago by its impact on Earth. No human being has seen this object. However, the impact crater is known and with computer models an approximate  size of this object has been established. The existence of this object is hardly questioned.  Its parameters have been determined to fit known facts.

Another example is given by the celestial spheres of  ancient Greek astronomy. The outermost crystal sphere contained all the fixed stars. It explained that the stars rotate with equal angular velocity around he immovable Earth. In this case, with more data the celestial spheres had to be abandoned in favor of a rotating Earth moving around the Sun, the stars being far away.

\subsection{The assumption of  a hot planet}
The asymmetry of the ice cover around the present North Pole \cite[fig.\thinspace 5]{Woelfli2011} is explained in an alternative theory
by a geographic shift of the poles. 
For this,   an object is postulated, which satisfies the following conditions:
\renewcommand{\labelenumi}{\Alph{enumi}:}
\begin{enumerate}
\item{The object is sufficiently heavy, so that during a near encounter with Earth it deforms Earth's globe slightly. This leads to the geographic motion of the Poles at the end of the Pleistocene.}
\item{The object does no more exist.}
\end{enumerate}

The discussion of the mechanism of a rapid geographic pole shift \cite{Woelfli2007}
shows, that the postulates A and B can be realized, provided 
 the following list of conditions holds. We admit that
 the processes are complex, while we could only use simple considerations, mostly order of magnitude estimates. More detailed investigations would be highly valuable. 

\renewcommand{\labelenumi}{\arabic{enumi}.}
\begin{enumerate}
\item{The object has to be sufficiently heavy, so that during a very close passage near the Earth its tidal force stretches the Earth in an oblique direction by about 1 permil.}
\item{The gyroscopic motion of the globe then produces the required geographic shift of the poles, while the rotation axis remains practically fixed in space.}
\item{The global deformation must relax in a time of the order of magnitude of Earth's precession time, which is about 400 days.}
\item{During the close passage the object must break to pieces  owing to the tidal force produced by Earth's mass.}
\item{The object has to be hot.}
\item{The object was in an extremely eccentric orbit near Earth's orbital plane for about 3\thinspace 000 ka. During this time it evaporated partially.}
\item{As a result, a disk shaped cloud near the invariant plane surrounded the Sun.}
\item{The cloud was a plasma consisting mostly of ions and electrons.}
\item{The cloud was generated by the evaporation from the hot planet. This evaporation was limited by the escape velocity. The cloud's density increased gradually.}
\item{When the density of particles in the cloud reached a critical value, collisions, which were mostly inelastic, set in. They lead to a collapse of the cloud. }
\item{When Earth's orbital plane was within the cloud, the cloud had a cooling effect. In a Dansgaard-Oeschger event the temperature dropped gradually as the cloud's density increased and the temperature rose rapidly due to the cloud's collapse \cite[fig.8]{Woelfli2011}.} 
\item{After the close passage, Earth's orbital plane made an angle of a few degrees with the cloud.}
\item{Then, the additional radiation scattered from the cloud had a warming effect. Therefore, the Dansgaard-Oeschger events were inverted: the temperature rose gradually and  dropped rapidly \cite[fig.11]{Woelfli2011}.}
\item{The Dansgaard-Oeschger temperature variations occurred also in the Antarctis with high temperature peaks at approximately the same times. However, the temperature differences were smaller and the warmings less abrupt \cite{Blunier}.}
\end{enumerate}

During the Holocene the cloud  was produced by the rapidly evaporating fractions of the hot planet. Therefore, it was especially dense.
During the cold 8.3 ka event  \cite[fig.11]{Woelfli2011}, Earth's orbit  must have been within this cloud.
The fact that afterwards the temperature was higher than before shows that  during this event the cloud's density continued to increase.

The gradual growth of the cloud followed by a rapid collapse lead to irregular climate periods of about 2 ka, which are too short to be caused by the Milankovitch perturbations of Earth's orbit. The fact that in the  warm period of the Holocene the temperature showed the reversed behavior 
 \cite[fig.11]{Woelfli2011}
compared to that of  the cold periods of the Pleistocene \cite[fig.\thinspace 8]{Woelfli2011} strongly favors the model in which a cloud decreases the temperature, when Earth moves within the cloud, and increases the temperature, when the orbital plane is outside. It may be difficult to find other climate relevant processes which invert their time behaviour depending on the temperature being higher or lower than at present.  The cloud obviously does not exist at present; this is in line with our relatively constant climate, in which human made perturbations become important. If the present were a warm period of the Ice Ages, such as the Eemian about 120 ka ago (which was warmer than the present), then 
the intrinsic climate fluctuations would be larger. If  at present we were still in the Ice Ages,  the important question would be, how human activities modify the intrinsic fluctuations, in particular, whether the gases CO$_2$ and CH$_4$ can avoid dropping into a cold period.

The gradual increase in density of the disk shaped cloud and its abrupt collapse produced variations in heat fluxes, which were equal on both hemispheres \cite{EPICA}. 
The fact, that the resulting temperature changes were different on the two hemispheres indicates that the thermohaline ÔÔbipolar see-sawÕÕ \cite{Broecker,StockerT} between northern and southern hemispheres was acting. This is a change of ocean currents driven by modifications of the water density due to  temperature and to salt content.  This has been studied extensively \cite{Stocker}. The heat transport by currents of the oceans and the atmosphere, in particular the Golf stream, exists. It is different on the two hemispheres.
 Again, this important part of the usual theory is real and necessary for the understanding of the climate events. However,
 in the alternative theory the changes  of the currents of oceans and atmosphere are triggered by variations of the heat flux from the disk shaped cloud.
 Possibly, Heinrich events \cite{StockerT} were triggered by earthquakes due to fairly close approaches of the hot planet.

Let us return to the main subject: the unusual object. Since in the alternative theory it was especially designed to explain the asymmetry of the glaciation, it will of course explain this feature.  
On the other hand, 
such an object has many other consequences. These were not foreseen when we  postulated this object.

 Velikovsky \cite{Velikovsky} interpreted old traditions, which clearly indicated an astronomical object. He concluded that Venus was the culprit of the disasters.  Einstein \cite{Einstein}, in an exchange of letters with Velikovsky, resumed his attitude writing:  
''To the point, I can say in short: catastrophes {\it yes,} Venus {\it no.}''
 Velikovski's choice of Venus had its reasons. Since Venus is close to the Sun, it  appears as an evening or morning star. The hot planet in its eccentric orbit  also appeared in the evening or morning, when it was particularly shiny or when it was menacing to come near the Earth. On the other hand,  it is inconceivable mechanically, how Venus could have changed into its present orbit far from the Earth.

The authors assume that the asymmetry of the ice shield requires a pole shift and that this has an astronomic cause. Their postulates A and B lead to the 
 conditions 1 to 14. The corresponding astronomical object  cannot be observed at present. 
 Its eccentric orbit limits its life to a few 1\thinspace 000 ka. Therefore, it must be a rare companion in any planetary system. This explains why such an object has never been observed.
 Originally, the object may have been a moon of Jupiter, which was expelled. In this case Jupiter's lunar system should still be in an unrelaxed state, which actually seems to be the case \cite{Maggie2013}. In principle,  the object might also have been a floating planet from interstellar space \cite{MOA}. However, in this  case a mechanism is needed, which brings its orbit into the ecliptic plane.
 
During the motion in an extremely eccentric orbit (in our estimates we often used the eccentricity $\epsilon = 0.97$), the hot planet slowly evaporated. This loss of material is limited by the escape velocity. The evaporated particles are single atoms or ions. Since the solar radiation disrupts molecules and ionizes atoms, a cloud consisting of multiply ionized atoms and of electrons is created. This disk shaped cloud is a plasma. It may have important properties (discharges, contraction of width of cloud, \ldots ), which, however, we do not understand . The cloud as a whole does not show the strong Milankovitch variations expected for an individual particle in the eccentric orbit \cite{Nufer}. 
It appears that
the cloud remains near the invariant plain, perpendicular to the angular momentum of the planetary system. 

At first sight
it seems that by introducing an object, such as a hot planet, one can also fit an elephant. After all, the object has been invented to explain certain observations.  The point is, that it is not easy to invent an object, which acted  during the Ice Ages and which has disappeared. Our estimates are rough. They indicate that the object must be at least Mars sized in order to  be able to stretch the Earth sufficiently (one per thousand) during the close approach. Furthermore, it must be at most Mars sized and hot so that it can be torn to pieces during this event. Fractions with their reduced escape velocity can evaporate during  the 10 ka of the Holocene. This theory leaves little space for arbitrary choices. It is a narrow window, which might close with other estimates than ours.

\subsection{Results of the alternative theory}

Some explanations of Ice Age phenomena are specific to this theory. In particular:
\renewcommand{\labelenumi}{\alph{enumi}:}
\begin{enumerate}

\item{The asymmetry of the glaciation  \cite[fig.5]{Woelfli2011} and the presence of mammoths in arctic regions of East Siberia are explained by a rapid geographic shift of the poles \cite[fig.6]{Woelfli2011} at the end of the Pleistocene.}

\item{The dominant 100 ka period is the Milankovitch period of the inclination, as has been pointed out by R.A. Muller and G.J. MacDonald \cite{Muller95,Muller97}. 
They postulated a disc shaped cloud around the Sun.
The hot planet creates this cloud.}

\item{The rapid Dansgaard-Oeschger temperature variations \cite[fig.8]{Woelfli2011} are caused by changes of the cloud, namely by the gradual increases of its density followed by abrupt collapses \cite{Woelfli2007}. In cold periods, i.e. when Earth's orbit is within the cloud, the temperature decreases gradually and increases abruptly.}

\item{At the end of the Pleistocene, the narrow passage of the hot planet led to a deflection of Earth's orbit by an angle of one or several degrees,.  Therefore, during the Holocene Earth's orbital plane was outside the cloud (excepting the 8.2 ka event) \cite[fig.11]{Woelfli2011}. }

\item{During the Holocene the Earth received additional radiation scattered from the cloud. This explains, why the temperature was higher than at present \cite[fig.10]{Woelfli2011}.}

\item{During the Holocene the temperature  increased gradually  and dropped abruptly \cite[fig.11]{Woelfli2011}. When the climate was warmer than at present, the Dansgaard-Oeschger events are reversed. Then, the cloud warms the Earth with scattered light, rather than  cooling it by shielding.}

\item{The observed duration of the Ice Ages  ($\approx$ 3\thinspace 000 ka ) corresponds to a typical number of trials  for a close passage  \cite[subsection 3.6 ]{Woelfli2011}}. The extremely eccentric orbit of the hot planet can exist during this period.

\end{enumerate}

The authors were looking for a model which solves item a.  This led to the conditions A and B. Since
the postulates A and B are strongly suggested by empirical data,  historically they have  been examined in detail. However, the conclusion was that they cannot be fulfilled.  The authors claim that with a hot planet in an extremely eccentric orbit and a relaxation time for the global deformation of the order of the precession time, the postulates A and B can be satisfied.  It turns out that this also explains phenomena  b to g. Clearly, this adds confidence in the model.

\subsection{Traditions}
To call the hot planet an unobserved object is not quite correct. It is unobserved at present. However, if it existed, it must have been a horrifying object in the life of our ancestors. Ancient populations made great efforts with astronomic observations. As late as in antique Greece planets were still gods. Now the positions of planets are irrelevant for daily life (except for persons who believe in astrology).   In \cite{Traditions} we have collected orally transmitted stories, which involve the hot planet. The subsection  \cite[2.7.1]{Traditions} contains a report by American indians of the cataclysmic passage of the hot planet. 

The appearance of the hot planet varied: when it returned from outside, it may have looked calm like the Japanese flag, while coming from near the Sun, it may have been menacing like a fire--spitting Chinese dragon.  When it no more existed, translations of traditions may have been misleading; it might have been called  ``morning star'' or ``evening star'' in view of the times, when  it could be threateningly visible, or even ``Sun'' in view of its heat and glow. 

I. Velikovsky \cite{Velikovsky} collected reports from many old cultures. Inspired by this, he gave a description of events, which can hardly be reconciled with  laws of nature.  Therefore, his work has been disregarded by many scientists. However, the access to traditions from various civilizations, which he uncovered, is of high value. In the field of recent paleoclimate, a better cooperation between humanities and sciences would be beneficial to both.

\end{document}